\documentclass[
    ,draft            
  ]
  {aipproc}

\layoutstyle{6x9}

\def\ra{\!\rightarrow\!}

\def\dsmunu{$D^+_s\ra\mu^+\nu$} 
\def\dstaunu{$D^+_s\ra\tau^+\nu$} 
\def\dsphipi{$D^+_s\ra\phi\pi^+$}
\def\btaunu{$B^+\ra\tau^+\nu$} 
 
\def\fds{$f^{}_{D^{}_s}$}

\def\dbar{\overline{D}{}^0}

\def\gevm{~GeV/$c^2$}
\def\mevm{~MeV/$c^2$}

\def\meve{~MeV}

\def\simge{\mathrel{%
   \rlap{\raise 0.511ex \hbox{$>$}}{\lower 0.511ex \hbox{$\sim$}}}}
\def\simle{\mathrel{
   \rlap{\raise 0.511ex \hbox{$<$}}{\lower 0.511ex \hbox{$\sim$}}}}

\begin{document}
\begin{flushright}
UCHEP-09-02
\end{flushright}
\vspace*{-0.34in}

\title{{\textbf{$B^+$} and \textbf{$D^+_s$} Decay Constants from Belle and Babar}}




\author{A. J. Schwartz}{
  address={Department of Physics, 
University of Cincinnati, P.O. Box 210011, Cincinnati, Ohio 45221}
}

\begin{abstract}
The Belle and Babar experiments have measured the branching
fractions for \btaunu\ and \dsmunu\ decays. From these 
measurements one can extract the $B^+$ and $D^+_s$ decay 
constants, which can be compared to lattice QCD calculations. 
For the $D^+_s$ decay constant, there is currently a $2.1\sigma$ 
difference between the calculated value and the measured value.
\end{abstract}

\maketitle

\section{Introduction}

\begin{figure}
  \includegraphics[height=.15\textheight]{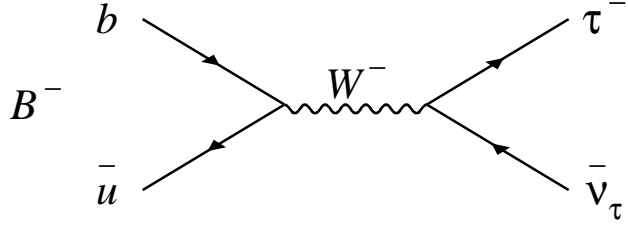}
\caption{Annihilation diagram for a heavy meson decaying to a lepton
and neutrino.}
\label{fig:diagram}
\end{figure}

Both the Belle~\cite{belle} and Babar~\cite{babar} experiments
have measured the branching fractions for \btaunu\ and \dsmunu\
decays~\cite{charge-conjugates}. These decays proceed via the 
annihilation diagram of Fig.~\ref{fig:diagram}. Within the 
Standard Model (SM), the predicted rates are
\begin{eqnarray}
{\cal B}(B^+\ra\tau^+\nu) & = & 
\tau^{}_{B^+}\,\frac{G^2_F}{8\pi}|V^{}_{ub}|^2 f^2_{B^+}\,m^2_{\tau}\,
m^{}_{B^+}\biggl( 1-\frac{m^2_{\tau}}{m^2_{B^+}}\biggr)^2  
\label{eqn:fb} \\
{\cal B}(D^+_s\ra\ell^+\nu) & = & 
\tau^{}_{D^{}_s}\,\frac{G^2_F}{8\pi}|V^{}_{cs}|^2 f^2_{D^{}_s}\,m^2_{\ell}\,
m^{}_{D^{}_s}\biggl( 1-\frac{m^2_{\ell}}{m^2_{D^{}_s}}\biggr)^2 \,.
\label{eqn:fds} 
\end{eqnarray}
For the $D^+_s$, all parameters on the right-hand-side 
except for \fds\ are well-known. The Cabibbo-Kobayashi-Maskawa (CKM) 
matrix element $|V^{}_{cs}|$ is well-constrained by a global fit to 
several experimental observables and unitarity of the CKM matrix.
Thus a measurement of ${\cal B}(D^+_s\ra\mu^+\nu)$
allows one to determine the decay constant~$f^{}_{D^{}_s}$.
This can be compared to QCD lattice calculations, which have
become relatively precise. For the $B^+$, the CKM matrix 
element $|V^{}_{ub}|$ is known to only 9\%; this is 
nonetheless more precise than measurements of
${\cal B}(B^+\ra\ell^+\nu)$ and allows one to extract~$f^{}_B$.

\section{Measurement of \textbf{\btaunu}}

\begin{figure}
\includegraphics[height=.50\textheight]{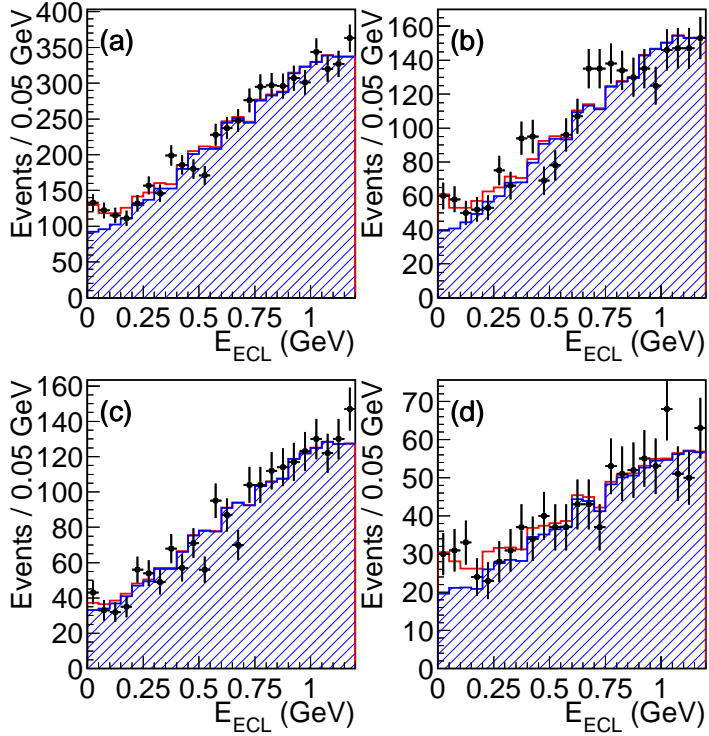}
\caption{
$E^{}_{ECL}$ distribution of data events (points) and 
fit projections for \btaunu, from Belle~\cite{btaunu_belle}. 
{\it (a)}\ all $\tau$ decay modes combined;
{\it (b)}\ $\tau^-\ra e^-\bar{\nu}^{}_e\,\nu^{}_\tau$; 
{\it (c)}\ $\tau^-\ra\mu^-\bar{\nu}^{}_\mu\,\nu^{}_\tau$; 
{\it (d)}\ $\tau^-\ra\pi^-\nu^{}_\tau$.
The open and hatched histograms correspond to signal and 
background, respectively.}
\label{fig:btaunu_belle}
\end{figure}

Belle has done two \btaunu\ analyses~\cite{btaunu_belle_old,btaunu_belle}; 
the most recent one used 605~fb$^{-1}$ of data and obtained 
evidence for a signal. This analysis employed a semileptonic tag:
one $B$ in an event is fully reconstructed as
$B^-\ra D^{(*)0}\ell^-\bar{\nu}$, where $D^{*0}\ra D^0\pi^0\!,\,D^0\gamma$
and $D^0\ra K^-\pi^+(\pi^0),\,K^-\pi^+\pi^-\pi^+$. 
The signal decays $\tau^+\ra\mu^+\nu_\mu\bar{\nu}^{}_\tau,\,
e^+\nu^{}_e\bar{\nu}^{}_\tau$
and $\tau^+\ra\pi^+\bar{\nu}^{}_\tau$ are then searched for by 
reconstructing a single track not associated with the tag side.
The signal yield is obtained by fitting the distribution 
of $E^{}_{\rm ECL}$, which is the energy sum of calorimeter 
clusters not associated with a charged track.
A peak near $E^{}_{\rm ECL}\!=\!0$ indicates a 
$\tau^+\ra\ell^+\nu^{}_\ell\bar{\nu}^{}_\tau$ or 
$\tau^+\ra\pi^+\bar{\nu}^{}_\tau$ 
decay. The main backgrounds are $b\ra c$ processes and 
$e^+e^-\ra q\bar{q}$ continuum events. The fit is unbinned 
and uses a likelihood function 
\begin{eqnarray}
{\cal L} & = & 
\frac{e^{-\sum_j n^{}_j}}{N!}\,
\prod_i \sum_j n^{}_j\,f^{}_j(E^{}_i)\,,
\end{eqnarray}
where $i$ runs over all events ($N$), $j$ runs over all signal
and background categories, $n^{}_j$ is the yield
of category $j$, and $f^{}_j$ is the probability density
function (PDF) for category $j$.
The branching fraction is calculated as
${\cal B}= n^{}_s/(2\cdot\varepsilon\cdot N^{}_{B^+B^-})$,
where $\varepsilon$ is the reconstruction efficiency as 
calculated from Monte Carlo (MC) simulation. 
The fit results are listed in Table~\ref{tab:btaunu_belle},
and the fit projections are shown in Fig.~\ref{fig:btaunu_belle}.
The systematic errors are dominated by uncertainty in the 
background PDF and the tag reconstruction efficiency. The
overall result is 
\begin{eqnarray}
{\cal B}(B^+\ra\tau^+\nu)\Bigr|^{}_{\rm Belle} & = & 
\Bigl(1.65\,^{+0.38}_{-0.37}\,^{+0.35}_{-0.37}\Bigr)\times 10^{-4}\,,
\label{eqn:btaunu_belle}
\end{eqnarray}
where the first error is statistical and the second is systematic.

\begin{table}
\renewcommand{\baselinestretch}{1.3}
\begin{tabular}{|l|ccc|}
\hline
Decay Mode & 
Signal Yield & 
$\varepsilon$ (\%) & 
${\cal B}\times 10^4$ \\ 
\hline
$\tau^-\ra e^-\bar{\nu}\nu^{}_\tau$ & $78\,^{+23}_{-22}$ & 0.059
& $2.02\,^{+0.59}_{-0.56}$ \\
$\tau^-\ra \mu^-\bar{\nu}\nu^{}_\tau$ & $15\,^{+18}_{-17}$ & 0.037
& $0.62\,^{+0.76}_{-0.71}$ \\
$\tau^-\ra\pi^-\nu^{}_\tau$ & $58\,^{+21}_{-20}$ & 0.047
& $1.88\,^{+0.70}_{-0.66}$ \\
\hline
Combined & 
$154\,^{+36}_{-35}$ & 0.143 & $1.65\,^{+0.38}_{-0.37}$ \\
\hline
\end{tabular}
\caption{Fit results for \btaunu, from Belle~\cite{btaunu_belle}.
The data sample corresponds to 605~fb$^{-1}$.}
\label{tab:btaunu_belle}
\end{table}

Babar has published two analyses of \btaunu\ decays: one using a 
semileptonic tag~\cite{btaunu_babar_semi} and the other using 
a hadronic tag~\cite{btaunu_babar_hadr}. Both analyses use
data samples consisting of $383\times 10^6$ $B\overline{B}$ pairs.
The former is similar to that used in the Belle analysis: the 
tag side is reconstructed 
as $B^-\ra D^{(*)0}\ell^-\bar{\nu}^{}_\ell$, where 
$D^{*0}\ra D^0\pi^0,\,D^0\gamma$ and 
$D^0\ra K^-\pi^+(\pi^0),\,K^-\pi^+\pi^-\pi^+\!,\,K^{}_S\,\pi^+\pi^-$.
Babar searches for
$\tau^+\ra\ell^+\nu^{}_\ell\bar{\nu}^{}_\tau$, 
$\tau^+\ra\pi^+\bar{\nu}^{}_\tau$, and also
$\tau^+\ra\pi^+\pi^0\bar{\nu}^{}_\tau$, where for the last mode 
the $\pi^+\pi^0$ mass is required to be near that of the $\rho^+$.
The signal is identified by plotting $E^{}_{\rm extra}$,
the energy sum of calorimeter clusters not associated with 
a charged track; a peak near zero indicates $\tau^+$ decay. 
The signal yield is obtained by 
counting events in a signal region, e.g., $E^{}_{\rm extra}<0.50$,
and subtracting off background as estimated from $E^{}_{\rm extra}$ 
sidebands. The number of events in the final sample is 245, the 
background estimate is $222\pm 13$, and the resulting 
branching fraction is 
$(0.9\,\pm 0.6{\rm\ (stat.)}\,\pm 0.1{\rm\ (syst.)})\times 10^{-4}$.

\begin{figure}
\includegraphics[height=.38\textheight]{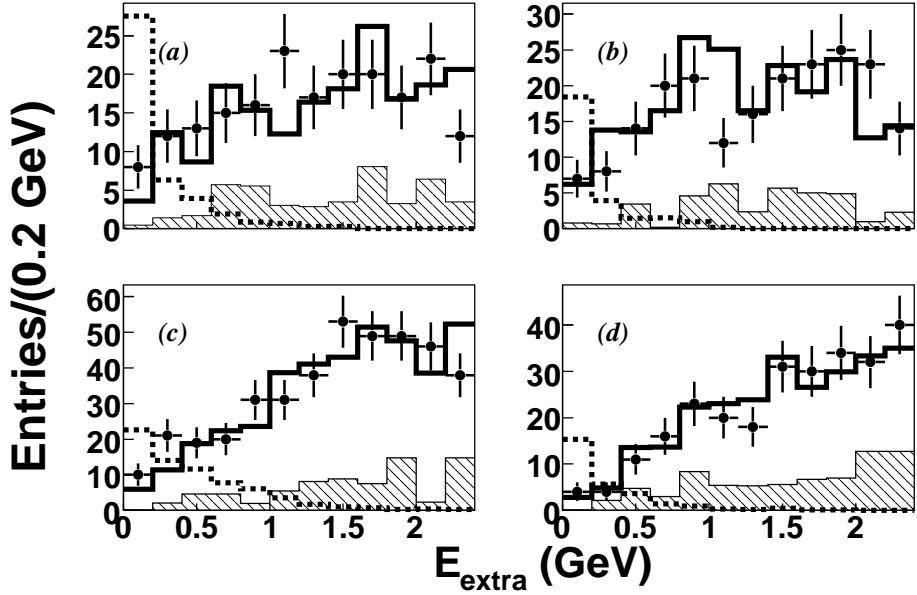}
\caption{
$E^{}_{\rm extra}$ distribution of data events (points) and 
fit projections for \btaunu, from Babar using a hadronic 
tag~\cite{btaunu_babar_hadr}. 
{\it (a)}\ $\tau^-\ra e^-\bar{\nu}^{}_e\,\nu^{}_\tau$; 
{\it (b)}\ $\tau^-\ra\mu^-\bar{\nu}^{}_\mu\,\nu^{}_\tau$; 
{\it (c)}\ $\tau^-\ra\pi^-\nu^{}_\tau$;
{\it (d)}\ $\tau^-\ra\pi^-\pi^0\nu^{}_\tau$.
The hatched histogram shows the combinatorial background
component, and, for comparison, the open histogram shows 
\btaunu\ signal for a branching fraction of 0.3\%.}
\label{fig:btaunu_babar}
\end{figure}

The Babar hadronic-tag analysis is more complicated.
The tagging side is reconstructed as $B^-\ra D^{(*)0}X^-$, where 
$X^-$ denotes a hadronic system of total charge $-1$ composed of 
$n^{}_1(\pi^\pm)$,
$n^{}_2(K^\pm)$,
$n^{}_3(K^0_S)$, and
$n^{}_4(\pi^0)$, where
$n^{}_1+n^{}_2\leq 5$, $n^{}_3\leq 2$, and $n^{}_4\leq 2$.
The $D^{(*)0}$ is reconstructed in the same channels as those 
used for the semileptonic analysis, as is the $\tau^+$ on the 
signal side. A background subtraction is done on the tag side.
The signal yield is obtained by counting events
in an $E^{}_{\rm extra}$ signal region and subtracting
off background as estimated from an $E^{}_{\rm extra}$ sideband.
There are 24 signal candidates and $14.3\pm3.0$ estimated
background events; the resulting branching fraction is 
$(1.8\,^{+0.9}_{-0.8}\,\pm 0.4\,\pm 0.2)\times 10^{-4}$,
where the first error is statistical, the second is due to
the background uncertainty, and the third is due to other
systematic sources. The data 
is shown in Fig.~\ref{fig:btaunu_babar} along with 
projections of the fit. This result is consistent with the 
semileptonic-tagged result; combining the two gives
\begin{eqnarray}
{\cal B}(B^+\ra\tau^+\nu)\Bigr|^{}_{\rm Babar} & = & 
\Bigl(1.2\,\pm 0.4{\rm\ (stat.)}\,\pm 0.3{\rm\ (bkg.)}\,\pm 0.2{\rm\ (syst.)}\Bigr)
\times 10^{-4}\,.
\end{eqnarray}
This is consistent with the Belle result, Eq.~(\ref{eqn:btaunu_belle}).

\section{Measurement of \dsmunu}

The Belle analysis of \dsmunu\ decays~\cite{dsmunu_belle}
uses 548~fb$^{-1}$ of data and searches for 
$e^+e^-\ra DKD^*_s\,n(\pi,\gamma)$, where the primary 
$D$ and $K$ can be charged or neutral; 
the $D^*_s$ is ``reconstructed'' (see below) via $D^+_s\gamma$;
$n(\pi,\gamma)\equiv X$ signifies any number of additional pions 
and up to one photon (from fragmentation);
the $D$ is reconstructed via $D\ra K\,n(\pi)$, where $n\!=\!1,2,3$;
and neutral kaons are reconstructed via $K^0_S\ra\pi^+\pi^-$.
If the primary $K$ is charged, both it and the $D$ must
have flavors opposite to that of the $D^+_s$;
these constitute a ``right-sign'' (RS) sample. If the
flavors are not both opposite, the event is categorized
as ``wrong-sign'' (WS) and used to parameterize the
background. The same classification applies to primary 
neutral $K$ events, except for these only the $D$ flavor 
must be opposite to that of the $D^+_s$ for the event to 
be classified as RS. 

\begin{figure}
\includegraphics[height=.40\textheight]{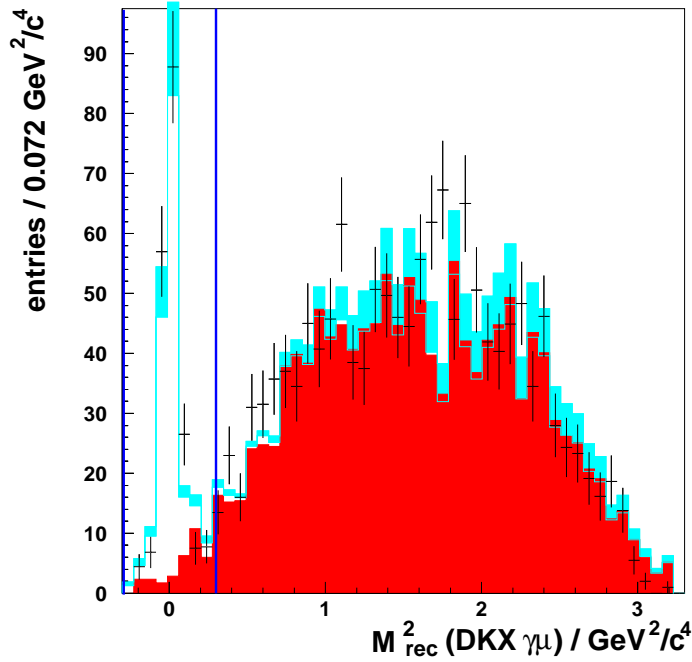}
\caption{Recoil mass distribution for $e^+e^-\ra DK\,X\gamma\mu^+$,
from Belle~\cite{dsmunu_belle}.}
\label{fig:widhalm1}
\end{figure}

The decay sequence is 
identified via a recoil mass technique. First, the
recoil mass of the $D$, $K$, and $X$ particles is
calculated and required to be within 150\mevm\ of
$M^{}_{D^*_s}$; then the $\gamma$ is included and
the recoil mass is required to be within 150\mevm\ of
$M^{}_{D^+_s}$; and finally, the $\mu^+$ is included 
and the recoil mass required to be within
0.55\gevm\ of zero. The final $DK\,X\gamma\mu^+$
recoil mass distribution is shown in Fig.~\ref{fig:widhalm1};
a sharp peak is observed near zero, indicating \dsmunu\ decay.

\begin{figure}
\vbox{
\vskip0.30in
\begin{center}
\includegraphics[height=.38\textheight]{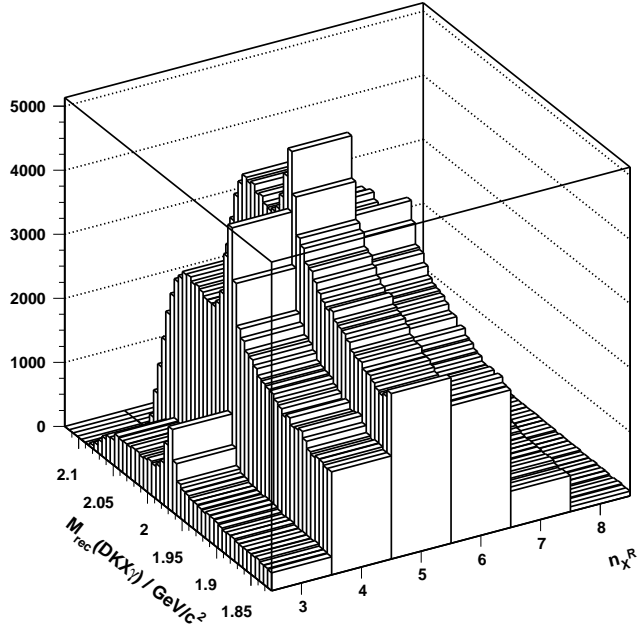}
\end{center}
}
\caption{Two-dimensional $DK\,X\gamma$ recoil mass
vs. $n^{R}_x$ distribution (see text), from Belle~\cite{dsmunu_belle}.}
\label{fig:widhalm2}
\end{figure}

The analysis is complicated by the fact that the recoil
mass technique is very sensitive to the number of tracks 
in an event and the track reconstruction efficiency, as 
all tracks must be reconstructed for the recoil mass to
be accurate. As it is difficult to simulate track 
multiplicity accurately due to uncertainties in quark 
fragmentation, the data is divided 
into bins of $n^{R}_x$, the number of ``primary particles'' 
reconstructed in an event. Here, a primary particle is one
that is not a daughter of any particle reconstructed in the
event. The minimum value for $n^{R}_x$ is three, corresponding
to $e^+e^-\ra D K D^*_s$ without any additional particles from
fragmentation. The data is then fit in two dimensions,
$DK\,X\gamma$ recoil mass vs. $n^{R}_x$
(see Fig.~\ref{fig:widhalm2}).
The signal PDF is obtained from MC and modeled
separately for different values of $n^{T}_x$, the
{\it true\/} number of primary particles in an event
($n^{T}_x$ can differ from $n^{R}_x$ due to particles being
lost or incorrectly assigned). 

The branching fraction is obtained from two fits:
the first fit uses the $DK\,X\gamma$ recoil mass spectrum
and yields the number of $D^+_s$ candidates; the result is 
$N^{}_{D^{}_s} = 32100\,\pm 870{\rm\ (stat)}\,\pm 1210{\rm\ (syst)}$.
For this fit the background shape is taken from the WS sample and 
the background levels floated in the fit.
The second fit uses the $DK\,X\gamma\mu^+$ recoil mass 
spectrum and yields the number of \dsmunu\ candidates; the result 
is $N^{}_{\mu\nu} = 169\,\pm 16{\rm\ (stat)}\,\pm 8{\rm\ (syst)}$.
For this fit the background shape is 
taken from a RS ``$D^+_s\ra e^+\nu$'' sample, i.e., all
selection criteria are the same except that an electron
candidate is required instead of a muon candidate. As
true $D^+_s\ra e^+\nu$ decays are suppressed by $\sim\!10^{-5}$,
this sample provides a good model of the \dsmunu\ background. 
The systematic errors listed are dominated by uncertainties in
the signal and background PDFs and are obtained by varying
the shapes of these PDFs.
The branching fraction is the ratio 
$N^{}_{\mu\nu}/N^{}_{D^{}_s}$, corrected for the ratio of
reconstruction efficiencies. The result is 
\begin{eqnarray}
{\cal B}(D^+_s\ra\mu^+\nu)\Bigl|^{}_{\rm Belle} 
& = & \Bigl(6.44\,\pm 0.76{\rm\ (stat.)}\,\pm 0.57{\rm\ (syst.)}\Bigr)\times 10^{-3}\,.
\end{eqnarray}

The Babar experiment searches for \dsmunu~\cite{dsmunu_babar}
using 230~fb$^{-1}$ of data by fully reconstructing a flavor-specific 
$D^{(*)-},\,\dbar$, or $D^-_s$ decay on the tagging side. Tag 
candidates are reconstructed in the following modes:
$\dbar\ra K^+\pi^-(\pi^0),\,K^+\pi^-\pi^+\pi^-$;
$D^-\ra K^+\pi^-\pi^-(\pi^0),\,K^0_S\,\pi^-(\pi^0),\,K^0_S\,\pi^-\pi^-\pi^+,\,
\,K^+K^-\pi^-,\,K^0_S\,K^-$; 
$D^-_s\ra K^0_S K^-,\,\phi\rho^-$; and 
$D^{*-}\ra\dbar\pi^-$ with
$\dbar\ra K^0_S\,\pi^+\pi^-(\pi^0),\,K^0_S\,K^+K^-,\,K^0_S\,\pi^0$.
An isolated $\mu^+$ track is required. The neutrino  
momentum is taken to be the missing momentum in the event: 
$\vec{p}^{}_\nu\equiv\vec{p}^{}_{e^+e^-}\!-\vec{p}^{}_{\rm rest}$.
A photon is required and paired with the $D^+_s$ candidate to make 
a $D^{*+}_s$, and the mass difference
$\Delta M\equiv M(\mu^+\nu\gamma) -M(\mu^+\nu)$ is calculated.

\begin{figure}
\includegraphics[height=.30\textheight]{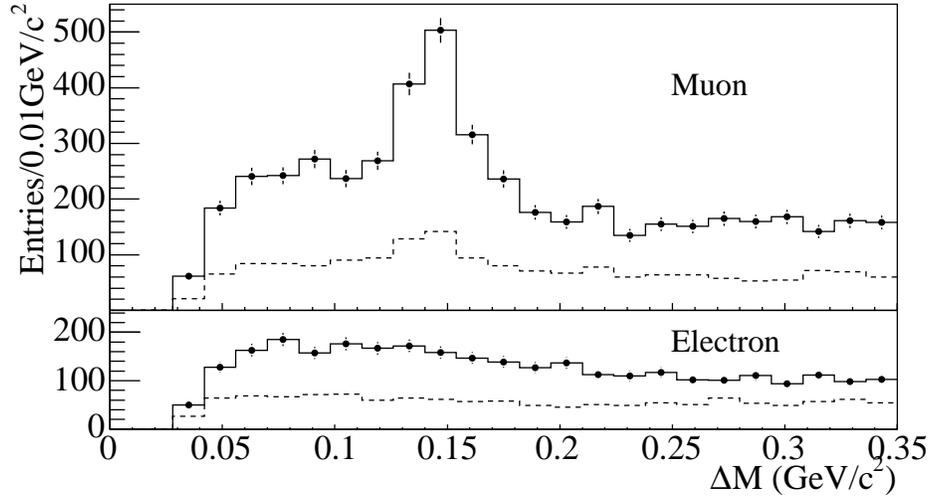}
\caption{$\Delta M$ spectra for the tag-side sideband region (dashed)
and tag-side signal region (solid) for $\mu^+$ (top) and $e^+$ (bottom)
samples, from Babar~\cite{dsmunu_babar}.}
\label{fig:dsmunu_babar1}
\end{figure}

\begin{figure}
\hbox{
\includegraphics[height=.18\textheight]{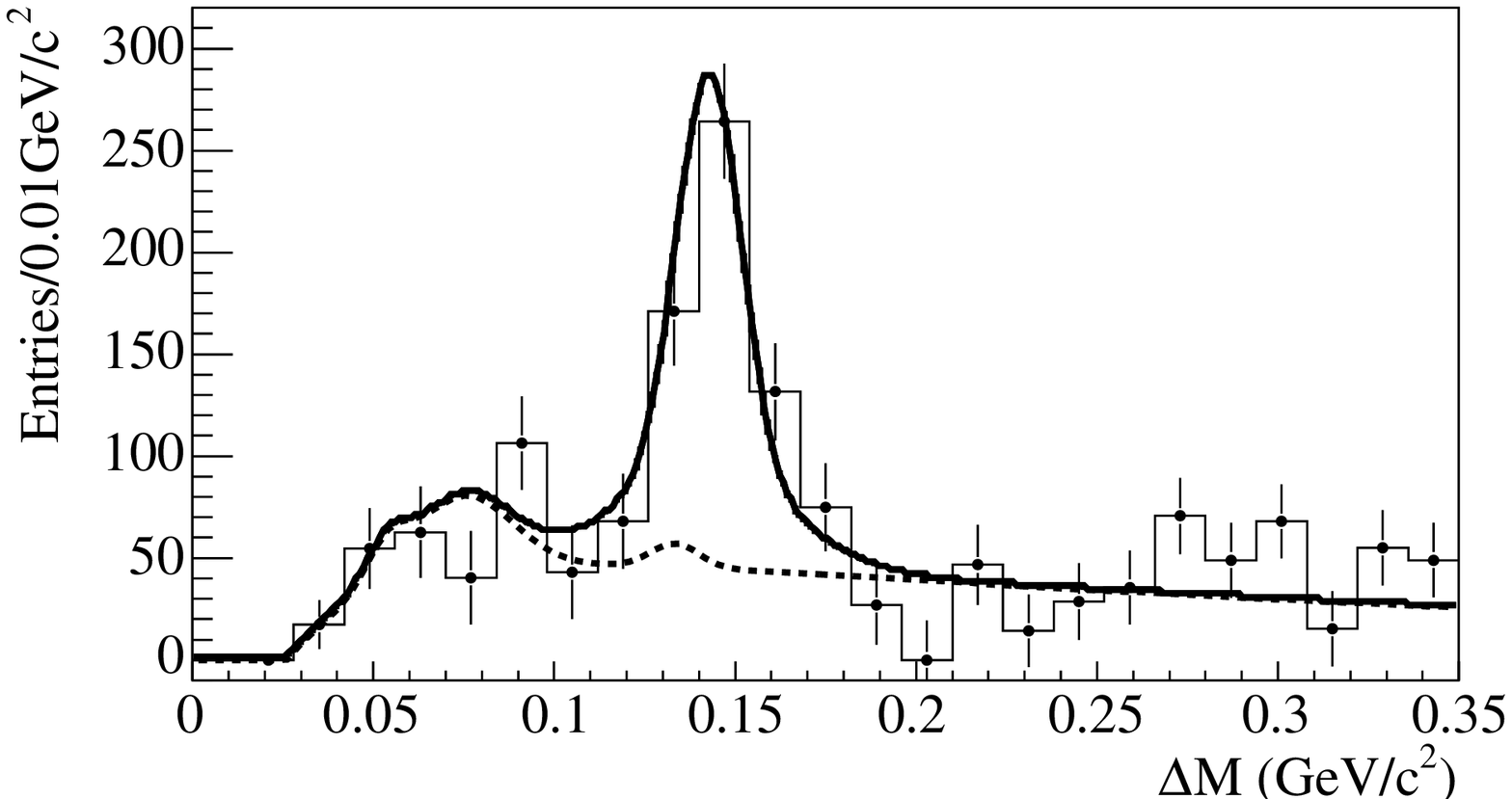}
\hskip0.02in
\includegraphics[height=.18\textheight]{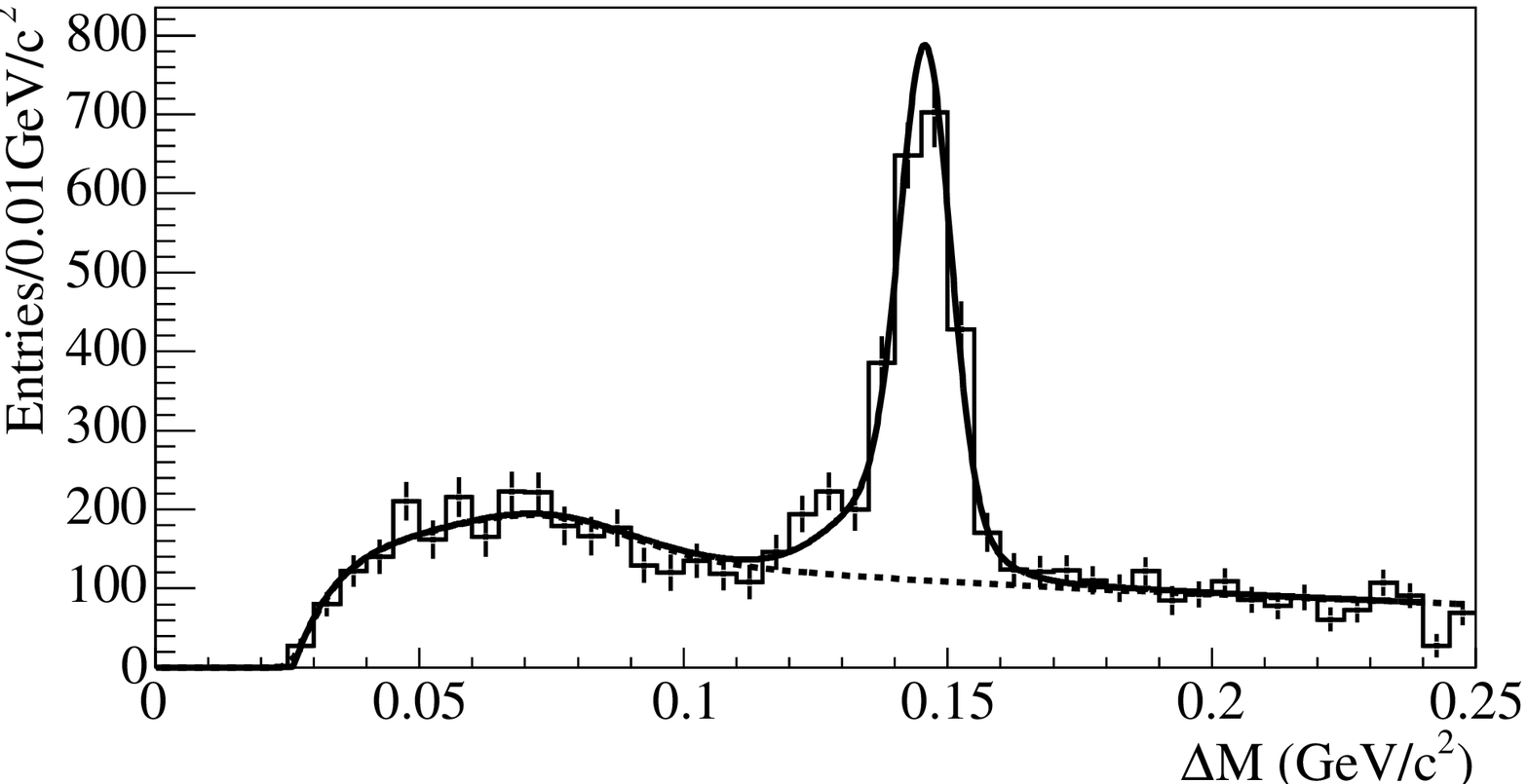}
}
\caption{Background-subtracted $\Delta M$ spectra (see text) and 
projections of the fit result, from Babar~\cite{dsmunu_babar}.
The left-most (right-most) distribution corresponds to 
\dsmunu\ (\dsphipi) candidates.
The dashed line is the background component, and the solid
line is the signal plus background components combined.}
\label{fig:dsmunu_babar2}
\end{figure}

The data is subsequently divided into four subsamples:
a tag-side mass sideband and a tag-side signal 
region for $\mu^+$ and $e^+$ candidates. For both 
lepton samples, the tag-side-sideband $\Delta M$ 
spectrum is subtracted from the tag-side-signal 
$\Delta M$ spectrum (Fig.~\ref{fig:dsmunu_babar1}), 
and then the sideband-subtracted
$e^+$ spectrum is subtracted from the 
sideband-subtracted $\mu^+$ spectrum. 
The final $\Delta M$ distribution 
(Fig.~\ref{fig:dsmunu_babar2}a) is
fit with signal and background PDFs; the signal
yield obtained is $N^{}_{\mu^+\nu} = 489\,\pm 55$ events.

To determine the branching fraction, the signal yield
is normalized to \dsphipi\ decays. Like
the signal mode, the $D^+_s$ candidate is required to
originate from $D^{*+}_s\ra D^+_s\gamma$.
The tag-side-sideband $\Delta M$ spectrum is subtracted
from the tag-side-signal $\Delta M$ spectrum, and the 
resulting spectrum is fit with signal and background PDFs
(Fig.~\ref{fig:dsmunu_babar2}b). The signal yield obtained 
is $N^{}_{\phi\pi^+} = 2093\,\pm 99$ events.
Dividing $N^{}_{\mu^+\nu}$ by $N^{}_{\phi\pi^+}$ 
and correcting for the ratio of reconstruction 
efficiencies gives
\begin{eqnarray}
\frac{\Gamma(D^+_s\ra\mu^+\nu)}{\Gamma(D^+_s\ra\phi\pi^+)}\Bigl|^{}_{\rm Babar}
& = & 0.143\,\pm 0.018{\rm\ (stat.)}\,\pm 0.006{\rm\ (syst.)}\,.
\label{eqn:dsratio_babar}
\end{eqnarray}

For this analysis, the $\phi$ is 
reconstructed via $\phi\ra K^+K^-$ with 
$|M^{}_{K^+K^-}-M^{}_\phi |\equiv\Delta M^{}_{KK}<5.5$\meve~\cite{joncoleman}.
Conveniently, CLEO has measured the branching fraction
${\cal B}(D^+_s\ra K^+K^-\pi^+)$ for $\Delta M^{}_{KK}=5{\rm\ MeV}$;
the result is $(1.69\,\pm 0.08\,\pm 0.06)\%$~\cite{cleo_dskkp}.
To multiply the two results together to obtain 
${\cal B}(D^+_s\ra\mu^+\nu)$ requires dividing 
Eq.~(\ref{eqn:dsratio_babar}) by ${\cal B}(\phi\ra K^+K^-) = 0.491$
and subtracting (in quadrature) the 1.2\% uncertainty in 
${\cal B}(\phi\ra K^+K^-)$ from the systematic error. In
addition, Babar has subtracted off a small amount of 
$D^+_s\ra f^{}_0(980)(K^+K^-)\pi^+$ background (48 events); 
as this process is included in the CLEO measurement, these 
events must be added back in to Babar's $\phi\pi^+$ yield. 
Thus the Babar result becomes
\begin{eqnarray}
\frac{\Gamma(D^+_s\ra\mu^+\nu)}{\Gamma(D^+_s\ra K^+K^-\pi^+)}
\biggr|^{}_{\Delta M^{}_{KK}=5.5{\rm\ MeV}} & = & 
0.285\,\pm 0.035{\rm\ (stat.)}\,\pm 0.011{\rm\ (syst.)}\,.
\label{eqn:ratio_adjusted}
\end{eqnarray}
Multiplying this by CLEO's measurement
gives
\begin{eqnarray}
{\cal B}(D^+_s\ra\mu^+\nu)\Bigl|^{}_{\rm Babar} &  = &
\Bigl(4.81\,\pm 0.63{\rm\ (stat.)}\,\pm 0.25{\rm\ (syst.)}\Bigr)\times 10^{-3}\,.
\end{eqnarray}


\section{Extraction of Decay Constants}

The Belle and Babar collaborations have used their 
measurements of ${\cal B}(B^+\ra\tau^+\nu)$ and 
Eq.~(\ref{eqn:fb}) to calculate the product of the $B$ decay constant
$f^{}_B$ and the CKM matrix element $|V^{}_{ub}|$. The results are
\begin{eqnarray*}
f^{}_B\times |V^{}_{ub}| \times 10^4 & = & 
\left\{
\begin{array}{l}
9.7\,\pm 1.1{\rm\ (stat.)}\,^{+1.0}_{-1.1}{\rm\ (syst.)}{\rm\ GeV\ \ 
(Belle)} \\ \\
7.2\,^{+2.0}_{-2.8}{\rm\ (stat.)}\,\pm 0.2{\rm\ (syst.)}{\rm\ GeV\ \ 
(Babar\ semileptonic)} \\ \\
10.1\,^{+2.3}_{-2.5}{\rm\ (stat.)}\,^{+1.2}_{-1.5}{\rm\ (syst.)}
{\rm\ GeV\ \ (Babar\ hadronic)}. 
\end{array}\right.
\end{eqnarray*}
Taking a weighted average gives
\begin{eqnarray}
f^{}_B\times |V^{}_{ub}|\,^{}_{\rm (Belle+Babar\ avg.)} & = & 
(9.2\pm 1.2)\times 10^{-4}{\rm\ \ GeV}\,,
\end{eqnarray}
and dividing by the Particle Data Group value
$|V^{}_{ub}| = (0.393\,\pm0.036)\%$~\cite{pdg_vub} gives
\begin{eqnarray}
f^{}_B\bigr|^{}_{\rm (Belle+Babar\ avg.)} & = & 233\pm 37~{\rm MeV}\,.
\end{eqnarray}
This value is $1\sigma$ higher than the most recent
lattice QCD results, that of the HPQCD collaboration
($190\,\pm 13$\meve~\cite{hpqcd_b}) and that of the
Fermilab/MILC collaboration ($195\,\pm 11$\meve~\cite{milc}).

\vskip0.20in

The Heavy Flavor Averaging Group (HFAG) has calculated 
a world average (WA) value for ${\cal B}(D^+_s\ra\mu^+\nu)$
and used this to determine a WA value for the $D^+_s$ decay 
constant \fds~\cite{hfag_ds}. This value can be compared to
recent lattice QCD calculations; a significant difference
could indicate new physics.
The WA value for \fds\ is obtained by inverting Eq.~(\ref{eqn:fds}):
\begin{eqnarray}
f^{}_{D^{}_s} & = & \frac{1}
{G^{}_F |V^{}_{cs}| m^{}_{\ell}
\biggl( 1-\frac{\displaystyle m^2_{\ell}}{\displaystyle m^2_{D^{}_s}}\biggr)}
\sqrt{\frac{8\pi\,{\cal B}(D^+_s\ra\ell^+\nu)}{m^{}_{D^{}_s} \tau^{}_{D^{}_s}}}\,,
\label{eqn:fds_inverted}
\end{eqnarray}
where, for ${\cal B}(D^+_s\ra\ell^+\nu)$, the WA value is inserted.
The error on \fds\ is calculated as follows:
values for variables on the right-hand-side of Eq.~(\ref{eqn:fds_inverted}) 
are sampled from Gaussian distributions having
means equal to the central values and standard deviations 
equal to their respective errors. The resulting values of 
\fds\ are plotted, and the distribution is fit to a
bifurcated Gaussian to obtain the $\pm 1\sigma$ errors.

\begin{figure}
\includegraphics[height=.53\textheight]{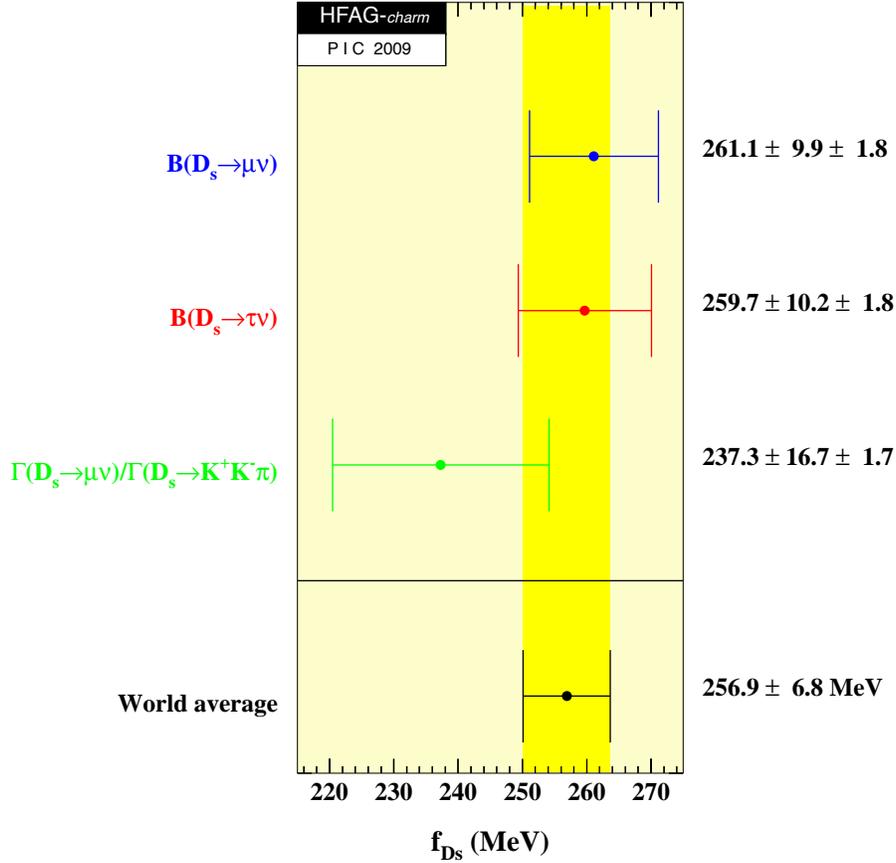}
\caption{Heavy Flavor Averaging Group (HFAG) WA value 
for \fds, from Ref.~\cite{hfag_ds}.
For each measurement, the first error listed is the total 
uncorrelated error, and the second error is the total
correlated error (mostly from $\tau^{}_{D^{}_s}$).}
\label{fig:hfag_fds}
\end{figure}

The results of this procedure are shown in Fig.~\ref{fig:hfag_fds}.
Also included are measurements of 
${\cal B}(D^+_s\ra\mu^+\nu)$~\cite{cleo_dsmunu} and 
${\cal B}(D^+_s\ra\tau^+\nu)$~\cite{cleo_dstaunu}
from CLEO. Thus there are three types of measurements:
\fds\ from the absolute \dsmunu\ branching fraction,
\fds\ from the absolute \dstaunu\ branching fraction, and
\fds\ from the $\Gamma(D^+_s\ra\mu^+\nu)/\Gamma(D^+_s\ra K^+K^-\pi^+)$
ratio. The overall WA value is obtained by averaging 
the three results, carefully accounting for 
correlations such as the input values for $|V^{}_{cs}|$ and 
$\tau^{}_{D^{}_s}$. The result is $256.9\,\pm 6.8$\meve. 
This value is higher than the two most precise lattice 
QCD results, that of the HPQCD ($241\,\pm 3$\meve~\cite{hpqcd_ds}) 
and Fermilab/MILC ($249\,\pm 11$\meve~\cite{milc}) collaborations.
The weighted average of the theory results is $241.5\,\pm 2.9$\meve,
which differs from the HFAG result by~$2.1\sigma$.

\section{Summary}

In summary, Belle has observed \btaunu\ with $3.8\sigma$ significance.
From the measured branching fraction they determine the product
$f^{}_B\times |V^{}_{ub}|$. Babar has observed \btaunu\ with $2.6\sigma$
significance and has also measured the branching fraction to determine
$f^{}_B\times |V^{}_{ub}|$. The results from the two experiments are
consistent; the weighted average has 13\% precision and is 
consistent with lattice QCD calculations.

For \dsmunu\ decays, Belle has observed this mode using a
recoil mass technique and has measured the branching fraction with
15\% precision. Babar has also observed this mode and has measured
the branching fraction relative to that for \dsphipi\ with
13\% precision. Dividing this by the branching fraction for 
$\phi\ra K^+K^-$ and including $D^+_s\ra f^{}_0(980)(K^+K^-)\pi^+$ 
decays allows one to multiply by CLEO's measurement of 
${\cal B}(D^+_s\ra K^+K^-\pi^+)$ to obtain
${\cal B}(D^+_s\ra\mu^+\nu)$. 
The Heavy Flavor Averaging Group has used the Belle and Babar 
measurements and also measurements from CLEO to calculate a 
world average value for \fds; the result is $256.9\,\pm 6.8$\meve. 
This value is $2.1\sigma$ higher than the average of two recent
lattice QCD calculations; the difference could indicate  
new physics.

\begin{theacknowledgments}

We thank the organizers of CIPANP 2009 for 
a stimulating scientific program and excellent hospitality. 
We thank Laurenz Widhalm, Yoshihide Sakai, and Andreas Kronfeld 
for reviewing this manuscript and suggesting many improvements.

\end{theacknowledgments}

\bibliographystyle{aipproc}   


\begin{thebibliography}{9}
\bibitem{charge-conjugates} 
Unless noted otherwise, charge-conjugate modes are implicitly included.
\end{thebibliography}


\end{document}